\documentclass[amsfonts,prd,aps]{revtex4}

\begin{document}

\title{Loop quantization of spherically symmetric midi-superspaces}

\author{ Miguel Campiglia$^{1}$, Rodolfo Gambini$^{1}$,
Jorge Pullin$^{2}$}
\affiliation {
1. Instituto de F\'{\i}sica, Facultad de Ciencias, 
Igu\'a 4225, esq. Mataojo, Montevideo, Uruguay. \\
2. Department of Physics and Astronomy, Louisiana State University,
Baton Rouge, LA 70803-4001}

\begin{abstract}
  We quantize the exterior of spherically symmetric vacuum space-times
  using a midi-superspace reduction within the Ashtekar new variables.
  Through a partial gauge fixing we eliminate the diffeomorphism
  constraint and are left with a Hamiltonian constraint that is first
  class. We complete the quantization in the loop representation.  We
  also use the model to discuss the issues that will arise in more
  general contexts in the ``uniform discretization'' approach to the
  dynamics.
\end{abstract}

\maketitle
\section{Introduction}

Loop quantum gravity has emerged in recent years as a significant
candidate for a theory of quantum gravity. See \cite{lqg} for recent
reviews. The theory has a mathematically rigorous basis for its
quantum kinematics \cite{kinematics}, which has also been proven to be
unique up to certain assumptions \cite{lost}.  The problem of the
dynamics of the full theory has remained unsettled. The origin of the
difficulties can be traced back to the kinematical space of states. In
this space, infinitesimal diffeomorphisms are not implemented as
operators. 
The issue is further compounded by the fact that
one cannot check the consistency of the quantum constraint algebra,
since the commutator of two Hamiltonians is proportional to an
infinitesimal diffeomorphism. Attempts to circumvent this problem,
like representing the Hamiltonian constraint as an operator
on states invariant under diffeomorphisms, are faced with other
difficulties (see \cite{thiemanngiesel} for a more thorough
discussion).

These problems have led several researchers to consider alternatives
to the usual Dirac approach to the problem of the dynamics. One of the
alternatives is the ``master constraint'' project of Thiemann and
collaborators \cite{phoenix} which has similarities to an earlier
proposal by Klauder \cite{klauder}.  Others consider the covariant
``spin foam'' approach as an alternative, since one may bypass the
construction of the canonical algebra entirely, at least in some
settings. Another approach is the one we have presented in recent
papers called ``uniform discretizations''
\cite{uniformprl,uniformprd}.

The spherically symmetric case has also specific problems  in the
traditional approach to loop quantum gravity. It has not been possible
up to now to find a particularization of the construction of Thiemann
for the Hamiltonian constraint to the spherically symmetric case that
has the appropriate algebra of constraints on the diffeomorphism
invariant space of states
\cite{boswi}.

In spite of these difficulties, the loop approach has been successful
in the context of homogeneous cosmologies, giving rise to ``loop
quantum cosmology'' \cite{lqc}. The reason why the approach works in
this context is that there is no issue with the algebra of
constraints: the diffeomorphism constraint is gauge fixed by the use
of manifestly homogeneous variables and there is only one Hamiltonian
constraint. We would like to show that a similar situation emerges in
the case of spherically symmetric space-times. We will show that one
can fix the diffeomorphism gauge in such a way that one is left with
only a set of Abelian Hamiltonian constraints.

Since this is our first approach to this problem, we will not explore
the more interesting possibilities of this model, for instance, what
happens to the singularity.  We will restrict our attention to the
exterior of the horizon.  We will also fix a gauge to eliminate the
diffeomorphism constraint for the sake of simplicity. The resulting
Hamiltonian constraint has nevertheless a non-trivial first class
algebra (with structure functions) in the continuum, but we will show
it can be Abelianized. We will show that the loop quantization can
be completed within the traditional Dirac quantization approach. In
particular we will construct the kinematical and physical Hilbert
spaces and we will show the quantization to be equivalent to the
one carried out in terms of traditional variables by Kucha\v{r}
\cite{kuchar}. We will discuss the implications for the elimination
of the singularity, though we will not work out the details in this
paper.

We would also like to discuss what lessons one can get from this model
for the ``uniform discretization'' approach to the dynamics of quantum
gravity. Since the uniform discretization approach is equivalent to
the Dirac quantization procedure if one has Abelian constraints, it
does not add anything new to the quantization of this model if one
chooses to Abelianize the constraints as we do.  We nevertheless would
like to discuss what would happen if one had chosen not to Abelianize
the constraints. The uniform discretization procedure can be applied,
but the treatment of the model becomes more involved, and the degree of
complexity increases with how much one chooses to ignore about the
particular details of the model.

The organization of this paper is  as follows. In section II we
set up the classical variables for spherically symmetric space-times
in the Ashtekar formulation; in section III we discuss the loop
quantization using the traditional Dirac procedure and in section
IV we discuss the use of uniform discretizations. We end with
a discussion.

\section{Spherically symmetric space-times}

We will use the Ashtekar new variables to describe the spherically
symmetric space-times. Previous work on this subject was done by
Bengtsson \cite{be}, Thiemann and Kastrup \cite{thiemannthesis} which
also Abelianize the constraints with the traditional (complex) Ashtekar
variables and with the more modern real connection variables
by Bojowald and Kastrup
\cite{boka}, and Bojowald and Swiderski \cite{boswi}  so we will not
repeat the full construction of spherically symmetric triads and
connections (the latter were also discussed in the context of
spherically symmetric Yang--Mills theory by Cordero and Teitelboim
\cite{cote}). Also, in the context of geometrodynamical variables,
gauge fixings in spherical symmetry were considered by \cite{husain}.

One assumes that the topology of the spatial manifold is
of the form  $\Sigma=R^+\times S^2$. We will choose a radial coordinate $x$
and study the theory in the range $[0,\infty]$. We will later assume that there
is a horizon at $x=0$, with appropriate boundary conditions as we discuss
below.

The invariant connection can be written 
as,
\begin{equation}
  A = A_x(x) \Lambda_3 dx + 
\left(A_1(x) \Lambda_1+ A_2(x) \Lambda_2\right)d\theta +
\left(\left(A_1(x) \Lambda_2- A_2(x) \Lambda_1\right)\sin \theta +
\Lambda_3 \cos \theta\right) d\varphi,
\end{equation}
where $A_x, A_1$ and $A_2$ are real arbitrary functions on $R^+$,
the $\Lambda_I$ are generators of $su(2)$, for instance $\Lambda_I = 
-i\sigma_I/2\equiv \tau_I$ where $\sigma_I$ are the Pauli matrices or
rigid rotations thereof. The invariant triad takes the form,
\begin{equation}
  E = E^x(x) \Lambda_3 \sin \theta {\partial \over \partial x} + 
\left(E^1(x) \Lambda_1 + E^2(x) \Lambda_2\right) \sin \theta {\partial \over 
\partial \theta} 
+
\left(E^1(x) \Lambda_2 - E^2(x) \Lambda_1\right) {\partial \over 
\partial \varphi},
\end{equation}
where again, $E^x, E^1$ and $E^2$ are functions on  $R^+$.  One
has the following canonical Poisson brackets for the symmetry reduced
variables,
\begin{eqnarray}
\left\{A_x(x), E^x(x')\right\}&=&2 \gamma G \delta(x-x'),\\
\left\{A_1(x), E^1(x')\right\}&=& \gamma G \delta(x-x'),\\
\left\{A_2(x), E^2(x')\right\}&=& \gamma G \delta(x-x'),
\end{eqnarray}
and all the other brackets vanish; $G$ is Newton's constant and 
$\gamma$ is the Immirzi parameter. 

Under coordinate transformations of the $x$ coordinate $x'(x)$, $A_x$ 
transforms like a scalar density $A'_x(x')=(\partial x/\partial x') A_x(x)$
whereas $E^x$ is a scalar. $A^1$ and $A^2$ are scalars and $E^1$ and
$E^2$ are scalar densities. 

There are three constraints. The first one is  a Gauss law,
\begin{equation}
G(\lambda) = \int dx \lambda \left((E^x)' + 2 A_1 E^2 -2 A_2 E^1\right),
\end{equation}
that generates $U(1)$ gauge transformations on the line. The second one
is the remnant of the  diffeomorphism constraint,
\begin{equation}
  D[N_x] =\int dx N_x \left( 2 (A_1)' E^1 + 2 (A_2)' E^2 -A_x (E^x)'\right),
\end{equation}
and finally the Hamiltonian constraint,
\begin{eqnarray}\label{23}
  H[N] &=& (2 G)^{-1} \int dx N(x)
\left( |E^x| \left[(E^1)^2+(E^2)^2\right]\right)^{-1/2}
\times \left( 2 E^x \left(E^1 A'_2-E^2 A'_1\right) + 2 A_x E^x\left (A_1 
E^1 +A_2 E^2\right)\right.\\
&&+\left(A_1^2+A_2^2-1\right)\left((E^1)^2+(E^2)^2\right)
-\left(1+\gamma^2\right)\left(2 K_x E^x \left(K_1 E^1+K_2 E^2\right)
+\left((K^1)^2+(K^2)^2\right)\left((E^1)^2+(E^2)^2\right)\right),
\nonumber
\end{eqnarray}
where $K_x$, $K_1$ and $K_2$ are the independent components of the
spherically symmetric curvature,
\begin{equation}
K=K_x \Lambda_3 dx+(K_1 \Lambda_1+K_2 \Lambda_2)d \theta+(K_1
\Lambda_2-K_2 \Lambda_1)\sin \theta d \varphi,
\end{equation}
which can be written in terms of the canonical variables $A$'s and
$E$'s.

It simplifies things if one introduces ``polar'' type coordinates in
the directions $1,2$. To do this we define,
\begin{eqnarray}
  A_1 &=& A_\varphi \cos \beta,\\
  A_2 &=&A_\varphi \sin \beta.
\end{eqnarray}
We introduce a canonical transformation from the variables
$A_1,A_2,E^1,E^2$ to the variables $A_\varphi,\beta,P^\varphi,P^\beta$
through the type III generating function $F=E^1 A_\varphi
\cos\beta+E^2 A_\varphi \sin\beta$. This leads to the above relation
between $A_{1,2}$ and $A_\varphi$ and $\beta$ and defines the
conjugate momenta,
\begin{eqnarray}
  P^\varphi &=& 2 E^1 \cos\beta +2 E^2 \sin\beta,\\
  P^\beta &=& -2 E^1 A_\varphi \sin \beta + 2 E^2 A_\varphi \cos\beta,
\end{eqnarray}
and introducing a new angular variable $\alpha$ via $E^1=E^\varphi
\cos\left(\alpha+\beta\right)$ and $E^2=E^\varphi 
\sin\left(\alpha+\beta\right)$,
we have that $P^\varphi=2 E^\varphi \cos \alpha$ and $P^\beta=2
E^\varphi A_\varphi \sin\alpha$, and we have rescaled $P^\varphi$ and
$P^\beta$ by a factor of two so one has the canonical Poisson bracket
relations,
\begin{eqnarray}
  \left\{A_\varphi(x),P^\varphi(x')\right\} &=& 2 \gamma G \delta(x-x'),\\
  \left\{\beta(x),P^\beta(x')\right\} &=& 2 \gamma G \delta(x-x'),
\end{eqnarray}
with the symplectic structure in $A_x,E^x$ unchanged from before.

In terms of these variables the Gauss law and diffeomorphism constraint
simplify,
\begin{eqnarray}
  G&=&P^\beta + (E^x)',\\
  D&=&P^\beta \beta'+P^\varphi A'_\varphi -(E^x)'A_x.
\end{eqnarray}

To write the Hamiltonian constraint it turns out that a further change
is desirable towards variables more closely connected with
the geometry. Let us start by identifying the metric constructed from the
densitized triads we have. To do this we first write the determinant of 
the metric ${\rm det}\, g = |E^x| (E^\varphi)^2 \sin^2 \theta$. 
The metric components are,
\begin{equation}
g_{xx}= \frac{(E^\varphi)^2}{|E^x|},\quad g_{\theta \theta} = |E^x|,\quad
g_{\varphi \varphi} = |E^x| \sin^2\theta.
\end{equation}
We consider generators of $su(2)$ rotated with
respect to the Cartesian basis. Given $\tan(\alpha+\beta)=-E^2/E^1$, we 
define,
\begin{eqnarray}
  \Lambda^\varphi_E &=& 
\Lambda_1 \cos(\alpha+\beta) +\Lambda_2 \sin(\alpha+\beta),\\
  \Lambda^\theta_E &=& 
-\Lambda_1 \sin(\alpha+\beta) +\Lambda_2 \cos(\alpha+\beta),
\end{eqnarray}
so the (undensitized) co-triad can be written as,
\begin{equation}
  e=e_x^3 \Lambda_3 dx +e_\theta^\theta \Lambda^\theta_E d\theta+
e_\varphi^\varphi \Lambda^\varphi_E d\varphi,
\end{equation}
where,
\begin{eqnarray}
  e_x^3&=& {E^\varphi \over {\rm sgn}(E^x)\sqrt{|E^x|}},\\
  e_\theta^\theta &=& \sqrt{|E^x|},\\
  e_\varphi^\varphi &=& \sqrt{|E^x|} \sin\theta.
\end{eqnarray}

From the co-triad we can compute the spin connection,
\begin{equation}
  \Gamma_x^3 =-(\alpha+\beta)',\quad
  \Gamma_\theta^\varphi ={(e^\varphi_\varphi)' \over e^3_x},\quad
  \Gamma_\varphi^\theta = -{(e^\varphi_\varphi)' \over e^3_x}\sin \theta,\quad
  \Gamma_\varphi^3 =\cos \theta,
\end{equation}
and then from $\gamma K_a^i= A_a^i-\Gamma_a^i$ we can compute the curvature
components
\begin{eqnarray}
\gamma K_x=A_x+(\alpha + \beta)', \\
\gamma K_1=A_1-\Gamma_\varphi \sin (\alpha + \beta), \\
\gamma K_2=A_2+\Gamma_\varphi \cos (\alpha + \beta),
\end{eqnarray}
where we define $\Gamma_\varphi$ as,
\begin{equation}
  \Gamma_\varphi \equiv -{(e^\varphi_\varphi)'\over e^3_x} = 
-{(E^x)'\over 2 E^\varphi},
\end{equation}
and also $\Gamma_x \equiv \Gamma^3_x$.

As was noted by Bojowald and Swiderski \cite{boswi}, in order to
introduce a representation in terms of holonomies one requires
certain falloff conditions, in particular $A_\varphi\to 0$ as
$x\to \infty$. This unfortunately is not true. To see it, note
that $\Gamma_\varphi$ is a ratio of two densities and therefore a
scalar. But asymptotically, $E^x \sim x^2$ and therefore $(E^x)'\sim2
x$ and $E^\varphi \sim x$ where $M$ is the mass of the
classical solution considered and therefore $\Gamma_\varphi \sim
1$, which implies that $A_\varphi$ will not tend to zero
asymptotically since $K_\varphi\to 0$ asymptotically and we have
the relation $(A_\varphi)^2=(\Gamma_\varphi)^2 + (\gamma
K_\varphi)^2$ ($K_\varphi=\sqrt{K_1^2+K_2^2})$. For further
discussion of asymptotic properties see Kucha\v{r} \cite{kuchar}.

To construct a connection with a good asymptotic behavior we 
consider the following canonical transformation,
\begin{eqnarray}
  A_\varphi \to \bar{A}_\varphi &=& 2 \cos \alpha A_\varphi,\\
  \beta \to \eta &=& \alpha+\beta,
\end{eqnarray}
with the following type II generating function
\begin{equation}
  F =P^\eta (\alpha+\beta) +2 E^\varphi A_\varphi \cos\alpha,
\end{equation}
which recovers the above transformation and introduces the 
canonical momenta $P^\eta$ and $E^\varphi$,
\begin{equation}
  P^\beta = P^\eta, \quad P^\varphi = 2 E^\varphi \cos \alpha,
\end{equation}
which leads us to consider the following canonical variables
$\bar{A}_\varphi,E^\varphi,\eta,P^\eta$, in terms of which the 
Gauss law and diffeomorphism constraint take a simple form,
\begin{eqnarray}
  G&=& P^\eta+(E^x)'\\
  D&=& P^\eta \eta' + E^\varphi\bar{A}_\varphi'-(E^x)' A_x.
\end{eqnarray}

Before continuing it is worthwhile noting that $E^x,E^\varphi$ are
``metric'' variables, in the sense that they are invariant under the
transformations generated by Gauss' law. $\bar{A}_\varphi$ and
$A_x+\eta'$ are also invariant and the latter 
is proportional to $K_x$. In order to clarify the meaning of
$\bar{A}_\varphi$ notice first that,
\begin{eqnarray}
  A_\varphi \Lambda_A^\varphi &\equiv& A_\varphi \cos \beta \Lambda_2
-A_\varphi \sin\beta \Lambda_1,\\
&=& A_1\Lambda_2 -A_2 \Lambda_1 = \Gamma_\varphi \Lambda^\theta_E+
\gamma K_\varphi \Lambda^\varphi_E,
\end{eqnarray}
and then note that $\Lambda^\theta_E$ and $\Lambda^\varphi_E$ are
orthogonal, i.e., ${\rm Tr}(\Lambda^\theta_E \Lambda^\varphi_E)=0$ and
also that ${\rm Tr}(\Lambda^\varphi_E \Lambda^\varphi_A)=\cos\alpha$.
Therefore $A_\varphi \cos\alpha = \gamma K_\varphi$ and it follows
that $\bar{A}_\varphi =2\gamma K_\varphi$. This automatically implies
that the connection has the right falloff condition since $K_\varphi\to
0$. We also have that,
\begin{equation}
  \gamma K_x = A_x -\Gamma_x =A_x + (\alpha+\beta)'=A_x+\eta',
\end{equation}
which confirms what had been noted, that $A_x+\eta'$
is a gauge invariant combination.


It is convenient to relate these variables of the Ashtekar formalism with
the canonical variables used for the spherical case by Kucha\v{r}. The
latter are given by,
\begin{equation}
  ds^2=\Lambda(x)^2 dx^2+R(x)^2 d\Omega^2,
\end{equation}
with $\Lambda> 0$ and $R>0$ and the latter is the curvature of the 2-spheres.
We then have,
\begin{eqnarray}
|E^x| &=&R^2\qquad {(E^\varphi)^2\over |E^x|} = \Lambda^2\\
K_{xx} &=& -\Lambda N^{-1}\left(\dot{\Lambda} 
-(N^x \Lambda)'\right)=-\Lambda K_x+2{\Lambda^2\over R}K_\varphi \\
K_{\theta\theta} &=& -N^{-1} R (\dot{R} -R' N^x) = -R K_\varphi \label{56}
\end{eqnarray}

One therefore has a canonical transformation between the pairs
$(\bar{A}_\varphi,E^\varphi,A_x+\eta',E^x)$ and
$(P_\Lambda,\Lambda,P_R,R)$ where,
\begin{eqnarray}
\Lambda &=& \frac{E^\varphi}{\sqrt{|E^x|}},\\
P_\Lambda &=& \sqrt{|E^x|} \frac{\bar{A}_\varphi}{2\gamma},\\
R &=& \sqrt{|E^x|},\\
P_R &=&  \sqrt{|E^x|} \frac{A_x+\eta'}{\gamma}+\frac{E^\varphi}{\sqrt{|E^x|}}
\frac{\bar{A}_\varphi}{2\gamma}.
\end{eqnarray}

The Hamiltonian constraint can be obtained from equation (\ref{23}), and
takes the form
\begin{eqnarray}
H&=& 
-\frac{E^\varphi}{2\sqrt{|E^x|}} 
-\frac{A_x \bar{A}_\varphi \sqrt{|E^x|}}{2\gamma^2}
-\frac{\bar{A}_\varphi^2 E^\varphi}{8\sqrt{|E^x|}\gamma^2}
+\frac{\left((E^x)'\right)^2}{8\sqrt{|E^x|}E^\varphi}\nonumber\\
&&-\frac{\sqrt{|E^x|} (E^x)' (E^\varphi)'}{2 (E^\varphi)^2}
-\frac{\bar{A}_\varphi \sqrt{|E^x|}\eta'}{2\gamma^2}
+\frac{\sqrt{|E^x|}(E^x)''}{2 E^\varphi}.
\end{eqnarray}

We will now fix the spatial diffeomorphism gauge freedom.  This
will simplify calculations considerably. We would like to have two
things: a) that the Gauss law remains as a constraint in order to have
usual loop representation techniques for quantization and b) that the
expression for the spatial volume retains a simple form since it 
plays an important role geometrically. We choose a gauge
that is suitable for the region exterior to a horizon, $E^x = (x+a)^2$
where $a$ is a positive constant given by $E^x|_{\rm Horizon}=a^2$, which
is equivalent to $R=x+a$ and $x=0$ corresponds to the horizon. This
gauge choice commutes with Gauss' law, which therefore remains a first
class constraint. Since we do not know the value of $E^x$ at the
horizon, $a$ is really a dynamical variable, its conjugate momentum is
related to $(A_x+\eta')|_{x=0}$. In that gauge the diffeomorphism
constraint takes the form,
\begin{equation}
  E^\varphi \bar{A}'_\varphi -(E^x)'(A_x+\eta') = E^\varphi \bar{A}'_\varphi
-2 (x+a) (A_x +\eta')=0,
\end{equation}
and it can be explicitly solved for $A_x+\eta'=E^\varphi
\bar{A}'_\varphi/(2(x+a))$.  This also determines strongly the value of the
canonical momentum of the variable $a$.

After solving for that variable, the Hamiltonian constraint takes the
form,
\begin{equation}\label{hamilclass}
  H=-{E^\varphi \over (x+a) \gamma^2}\left({\bar{A}^2_\varphi (x+a)\over 8}\right)'
-{E^\varphi \over 2 (x+a)}+ {3 (x+a) \over 2 E^\varphi} 
+ (x+a)^2 \left({1 \over E^\varphi}\right)'=0.
\end{equation}
The constraint only depends on the canonical pair
$\bar{A}_\varphi(x),E^\varphi(x)$ and $a$. The constraints remaining
after eliminating the diffeomorphism constraint are first class, but
have an algebra with structure functions,
\begin{eqnarray}
  \left\{H(x),H(y)\right\} &=& 
\left( {\bar{A}_\varphi(y)\over 2 \gamma} 
H(y)\right)' \delta(x-y)
-{\bar{A}_\varphi(y)\over  \gamma} 
H(y) \delta_{,x}(x-y),\\
  \left\{G(x),H(y)\right\} &=& 0,\\
  \left\{G(x),G(y)\right\} &=& 0.
\end{eqnarray}
This reduced system after eliminating the diffeomorphism constraint
is the one we will use as starting point for the quantization.

Our gauge choice is suitable for describing the exterior of a black
hole.  The initial data for the system is given on a spatial surface
extending from the Schwarzschild radius to infinity and it determines
entirely the spacetime bounded by the horizon and null infinity, since
no entering data is given on these two surfaces. The boundary
conditions on the horizon correspond to having the
$g_{xx}=(E^\varphi)^2/(x+a)^2$ component be singular,
$1/E^\varphi|_{x=0}=0$ and $\bar{A}_\varphi|_{x=0}=0$ (as discussed by
Bojowald and Swiderski \cite{bosw2}) which corresponds to the isolated
horizon boundary condition \cite{boswi}.  For the falloff conditions
at infinity we have \cite{kuchar},
\begin{eqnarray}
  E^\varphi &=& x + M(t) +O(x^{-\epsilon}),\\
  \bar{A}_\varphi &=& O(x^{-(1+\epsilon)}),
\end{eqnarray}
where $M(t)$ is a function of $t$ and the above expressions hold in the 
case that the value of $x\gg M(t)$. One also has that the shift
vanishes asymptotically $N^x=O(x^{-\epsilon})$ and the lapse $N=N+
O(x^{-\epsilon})$.

In order for the variations of the dynamical variables to preserve the
falloff conditions one has to add boundary terms to the action. Given,
\begin{equation}
  S(\bar{A}_\varphi,E^\varphi,N,a) = \lim_{x_+\to \infty} 
\int dt \int_0^{x_+} dx 
\left(E^\varphi \dot{\bar A}_\varphi - N H(x)\right).
\end{equation}
The problematic term is the one stemming from the variation of $E^\varphi$
in the term $(x+a)^2(1/E^\varphi)'$ in the Hamiltonian. Noting that 
$\delta E^\varphi_+=\delta M_+$ and that in problematic term one has,
\begin{equation}
  \int dx N (x+a)^2 \left(({E^\varphi})^{-2}(\delta E^\varphi)'\right).
\end{equation}
This term requires that one integrate by parts. This produces an extra
term of the form $N(x_+)\delta M_+$ and it would require that the lapse
vanish at infinity. This led Kucha\v{r} to propose adding a term
at infinity of the form $-\int dt M_+ \dot{\tau}_+$ with $\tau_+$ a
new dynamical variable such that its variation yields $\dot{M}_+=0$
and variation with respect to $E^\varphi_+$ is now well defined and
yields $\dot{\tau}_+=N_+$. The resulting action is,
\begin{equation}
  S(\bar{A}_\varphi,E^\varphi,N,a,\tau) = \lim_{x_+\to \infty} 
\int dt \int_0^{x_+} dx 
\left(E^\varphi \dot{A}_\varphi - N H(x)\right)-\int dt M_+ \dot{\tau}_+.
\end{equation}

Through the partial gauge fixing we are now left with a model that is
still a midi-superspace but that has only one constraint, the Hamiltonian,
with a non-trivial first class algebra.

\section{Quantization}

Our objective is to use the spherically symmetric model to explore
the properties of the loop quantization approach in a case
with field theoretic variables. Although the model in the end
has a finite number of degrees of freedom, this is achieved in a
non-trivial way  through the imposition of the Hamiltonian
constraint, all the time treating the model as a field theory. 

We will for simplicity concentrate on the exterior of the black hole
spacetime. We will therefore not concern ourselves in this first
approach on issues like how the use of loop variables may eliminate
the singularity, etc. The issues we are interested in probing are: the
role of the Hamiltonian constraint in generating the evolution, seeing
if the theory is coordinate invariant after quantization and
discretization and how the loop treatment compares with the continuum
quantization and seeing if one can avoid dealing with anomalies in the
quantum constraint algebra.  

\subsection{Traditional canonical quantization}

In order to compare with the other quantization approaches, we briefly
review the traditional canonical quantization of this model, first
discussed by Kucha\v{r} \cite{kuchar} (see this paper for earlier
reference on this subject).  In the region $x>0$ exterior
to the horizon, one can fix a gauge globally. For instance
$\bar{A}_\varphi=2\gamma K_\varphi=0$.  The constraints $\bar{A}_\varphi=0$ and
$H(x)=0$ are second class, since their Poisson bracket is
non-vanishing (not even weakly).  To quantize one has to impose these
constraints strongly. In the action the only remaining contribution
after this gauge fixing is an asymptotic one,
\begin{equation}
  S = \int dt \dot{\tau} M_+.
\end{equation}
From here one gets canonically conjugate variables $\tau,
P^\tau=-M_+$. $\tau$ is the proper time that determines the
position of the  spatial hypersurfaces  of vanishing extrinsic
curvature (usual Schwarzschild slicings). For these surfaces one
has $\bar{A}_\varphi=0$ and the quantity $E^\varphi$ 
is the solution of,
\begin{equation}
  H=0=-{(x+a)^2 E'_\varphi\over (E^\varphi)^2} + {3 (x+a) \over 2 E^\varphi}
-{E^\varphi \over 2 (x+a)},
\end{equation}
which yields as solution
\begin{equation}
E^\varphi=\frac{(x+a)}{\sqrt{1-\frac{a}{(x+a)}}}
=\frac{R}{\sqrt{1-\frac{a}{R}}}
\end{equation}
and one can identify by expanding asymptotically that $M=a/2$ and
recalling that $g_{xx}=(E^\varphi)^2/E^x$ one obtains the usual form
of the Schwarzschild metric with $M$ the mass of the space-time.

The quantization is straightforward, since the only remaining canonical
variables are $M$ and $\tau$. These variables have no dynamics. One has
$\hat{A}_\varphi=0$, 
$\hat{E}^\varphi=(x+2\hat{M})/\sqrt{1 -2 \hat{M}/(x+2\hat{M})}$ and
one can for instance introduce an eigenbasis of the mass operator,
$\hat{M} \phi(m)=m \phi(m)$. 

The operators associated to the other components of the metric can be
determined easily. The lapse is determined by the equation fixing the
preservation in time of the gauge condition, 
$\dot{\bar{A}}_\varphi = \{\bar{A}_\varphi,
\int N(x) H(x) dx\}=0$ with solution $N=\sqrt{1-a/(x+a)}$.
The shift is determined by noting that if $\bar{A}_\varphi=0$
then $K_\varphi=0$ and therefore equation (\ref{56}) implies, given
that $\dot{R}=0$, that $N^x=0$
One then
 has $g_{00}=N^2$ and this leads to the usual Schwarzschild solution:
 $\hat{g}_{00}=(1 -2\hat{M}/R)$.

Different choices of gauge lead to quantum theories in different
coordinate systems.  In the  gauge chosen above $\bar{A}_\varphi=0$.
For an more general gauge with arbitrary $\bar{A}_\varphi$ one has
 that
\begin{equation}\label{77}
E^\varphi=(x+2M)
\left[1-\frac{2M}{x+2M}+\frac{\bar{A}_\varphi^2}{4\gamma^2}\right]^
{-\frac{1}{2}}.
\end{equation}

For instance in the gauge
$\bar{A}_\varphi=\gamma \kappa x/(x+2M)^2$, with $\kappa$ a constant, 
and which satisfies the boundary
condition at the horizon and at infinity, one has that,
\begin{equation}\label{metricgauge1}
  E^\varphi = { x +2M\over \sqrt{1 -{2M\over x+2M}
+\frac{\kappa x}{4(x+2M)}}}
\end{equation}
and the lapse and shift are both non-vanishing,
\begin{eqnarray}\label{metricgauge2}
N&=&\sqrt{1-\frac{2M}{x+2M}+\frac{k^2}
{4(x+2M)^2}\left(1-\frac{2M}{x+2M}\right)^2},\\
N_x&=&\frac{N A_\varphi}{2\gamma}.\label{metricgauge3}
\end{eqnarray}

In the end all the quantizations are equivalent since in essence
we are dealing with a mechanical system and the choices of coordinates
are just different choices of quantities that one computes for the
mechanical system, and one has only two observables, the mass and the 
proper time at infinity. 

\subsection{Loop quantization}

In the loop representation the fundamental operators are 
associated with holonomies. This requires re-casting the theory of
interest in terms of such variables.
This usually involves taking limits of holonomies around loops of
vanishing area, for instance, to represent the curvature. Moreover,
the operators we wish to consider need to be regularized. This is
usually achieved by discretizing the theory on a lattice. Therefore
the usual loop treatment of theories is a natural framework in which 
to apply the ``uniform discretization'' technique.

We will proceed as follows:

1) We will identify the space of states of the loop representation for
the spherically symmetric case and the operators that are well defined
in this space.

2) We will particularize the space of states to the gauge we are
considering in this paper, where one is left with only the Hamiltonian
constraint.

3) We will note that one can Abelianize the constraint and, upon
discretization, the resulting discrete theory can be treated using the
traditional Dirac quantization procedure (in the case of Abelian
constraints it is known to be equivalent to the uniform discretization
procedure).

\subsection{The space of states 
in the loop representation for spherical symmetry}

The quantization in the loop representation is based on cylindrical
functions that depend on the connection through ``open
holonomies''. The latter are associated with graphs on the spatial
manifold (in this case the one dimensional line) composed of a set of
edges without intersections $g=\bigcup_i e_i$ where $g$ is the graph
and $e_i$ are the edges. The edges' vertices form a set $V(g)$
composed by all the endpoints of $e_i$. The $\bar{A}_\varphi$ correspond to
directions transverse to the radial one and are represented in the
loop representation via ``point holonomies'' $\exp(i \mu_v
\bar{A}_\varphi(x))=\exp(2 i \gamma \mu_v K_\varphi)$ at each vertex $v$
with $\mu\in \mathbb{R}$ as is usually done for scalar fields
\cite{bojowald}.  The variable $\eta$ is an angle and the corresponding point
holonomy is given by $\exp\left(i n_v
\eta(v)\right)$ with $n_v\in\mathbb{Z}$.  Before imposing gauge
invariance under the transformations generated by Gauss' law the spin
network states in a basis of spherically symmetric connections are,
\begin{equation}
  T_{g,\vec{k},\vec{n},\vec{\mu}}(A)=\prod_{e \in g} \exp\left(
{i \over 2} k_e \int_e {A}_x dx\right) \prod_{v\in V(g)} \exp\left(
i\mu_v \gamma K_\varphi(v)\right) \exp\left(i n_v \eta(v)\right)
\end{equation}
where $k_e \in \mathbb{Z}$ is the multiplicity of the loop. If we
recall that the Gauss law is $P^\eta +(E^x)'=0$ and that $\bar{A}_\varphi$
and ${A}_x+\eta'$ are gauge invariant, one has that the gauge invariant
spin networks for the spherically symmetric case are,
\begin{equation}
  T_{g,\vec{k},\vec{\mu}}(A) = \prod_{e \in g} \exp\left({i \over 2} k_e \int_e
\left(A_x+\eta'\right)dx\right) \prod_{v\in V(g)}
\exp\left(2 i \mu_v \gamma K_\varphi(v)\right).
\end{equation}

In the full three dimensional case the Hamiltonian constraint is
written in terms of holonomies along small closed paths. One can do
something similar here for the holonomies along the radial direction,
\begin{equation}
  \exp\left( i \int_{v_o}^{v_f} A_x dx\right)=1+i \epsilon A_x(v) +O(\epsilon^2),
\end{equation}
where $\epsilon=v_f-v_o$ is the distance between two vertices connected
by an edge. For the directions where we have homogeneity we cannot use
this argument since we do not have edges that we can make ``small''. 
In that case we can seek an alternative: to expand the ``point holonomies''
in the limit in which $\rho$ is small,
\begin{equation}
  \exp\left(i\gamma \rho K_\varphi(v)\right)=
1 + i \gamma \rho K_\varphi +O(\rho^2)\ldots
\end{equation}
It should be noted that quantum mechanically the limits $\epsilon\to
0$ and $\rho\to 0$ end up not being well defined and the resulting
quantum theories therefore approximate the classical theory well in
regions where the relevant variables in the exponent are small in
order to make the above expansions accurate.

At this point it is worthwhile mentioning that in the full 3D case, in
the loop representation, one represents the Hamiltonian constraint on
the space of diffeomorphism invariant states. Here we will not proceed
in such a way. To begin with, since we already fixed gauge for the
diffeomorphism constraint, there is no sense in representing the
Hamiltonian in a space of diffeomorphism invariant states.  Moreover,
if one were to try to mimic the construction that is done in the 3D
case (without fixing the gauge for diffeomorphisms), in the spherical
case one would encounter and additional difficulty: since there is no
notion of ``planar vertices'' the constraints will not close the
appropriate algebra even if represented on diffeomorphism invariant
states, at least with the usual treatment, as discussed in
\cite{boswi}. Therefore even if one started with diffeomorphism
invariant states the theory would not be invariant under space-time
diffeomorphisms.

\subsection{Loop representation of the partially gauge fixed theory}

We will work with a space where the spatial coordinates have been
gauge fixed in the way we described in the classical section, and
therefore the diffeomorphism constraint is not present. The relevant
canonical variables are $\bar{A}_\varphi,E^\varphi$ (we will usually
substitute $\bar{A}_\varphi$ for $K_\varphi$ since they are simply related)
and the asymptotic variables $M=a/2,\tau$.


We will work in a one dimensional lattice $L$ with points $0,\ldots x_N$ 
in $\mathbb{R}^+$ with spacing $x_{n+1}-x_{n}=\epsilon(n)$,
and we include the possibility of unequal spacing since we will see
it may be useful in certain cases. Strictly speaking, for the example
we are considering, there is no need to discretize and one could 
work directly in the continuum. We will work in a discretized lattice
to be able to make better contact later with the ``uniform discretization''
approach and also for comparison with other situations. We will see that in
the end we can correctly remove the lattice regulator and produce a
continuum theory.

The Hilbert space in the bulk is given by,
\begin{equation}
{\cal H} = L^2\left(\otimes_N R_{\rm Bohr},\otimes_N d \mu_0\right)
\end{equation}
where $R_{\rm Bohr}$ is the Bohr compactification of the real line and 
$\mu_0$ is the measure of integration we discuss below. In this space one
can introduce a basis,
\begin{equation}
T_{g,\vec{\mu}}[K_\varphi]=\prod_{v\in V(g)} 
\exp\left( i\mu_v \gamma K_\varphi(v)\right)
\end{equation}
with $\mu_v\in\mathbb{R}$, $g$ the graph, $V(g)$ vertices of $g$ in
$L$ and the notation $\vec{\mu}$ denotes
$(\mu_{v_1},\ldots,\mu_{v_p})$ with $p$ the number of vertices in the
graph. In the gauge fixed case we are considering a graph is just a
collection of vertices. We will now introduce a notation more adapted
to the uniform discretization setting by labeling the points in $L$
with an index $m,n,\ldots$. We would then have $K_{\varphi,n}=
K_\varphi(x_n)$. We have also rescaled $E^\varphi_n$ in such a way
that $\{K_{\varphi,m},E^\varphi_n\}=G \delta_{m,n}$ with $G$ Newton's
constant and we choose units where $\hbar=c=1$ and therefore 
$G=\ell_{\rm Planck}^2$.  
One then has the quantum representation, acting on states
$\Psi[K_{\varphi,m}]=\langle K_{\varphi,m}|\Psi\rangle$,
\begin{equation}
  \hat{E}^\varphi_m = -i \ell_{\rm Planck}^2 \frac{\partial}
{\partial K_{\varphi,m}},
\end{equation}
and therefore,
\begin{equation}
  \hat{E}^\varphi_m T_{g,\vec{\mu}} =\sum_{v\in V(g)} \mu_m
\gamma \ell_{\rm Planck}^2 \delta_{m,n(v)} T_{g,\vec{\mu}},
\end{equation}
where the $\delta$ is a Kronecker delta and $n(v)$ is the position on the 
lattice $L$ of the vertex $v$ of the spin network.

Given an interval $I$ in $L$ the volume of the corresponding ``shell''
is given classically by,
\begin{equation}
  V(I) = 4\pi \sum_{m\in I} |E^\varphi_m| (x_m+a),
\end{equation}
and as a quantum operator,
\begin{equation}
  \hat{V}(I) T_{g,\vec{\mu}} = \sum_{v \in I} 4\pi |\mu_v| (x_v+\hat{a}) \gamma
\ell_{\rm Planck}^2 T_{g,\vec{\mu}}. 
\end{equation}
We can introduce a basis of loop states $|g,\vec{\mu}>$,
\begin{equation}
  \langle K_{\varphi,m}|g,\vec{\mu}\rangle = T_{g,\vec{\mu}}[{K}],
\end{equation}
and the Bohr measure guarantees that,
\begin{equation}
  \langle g,\vec{\mu}|g',\vec{\mu}'\rangle
= \delta_{g,g'}\delta_{\vec{\mu},\vec{\mu}'}.
\end{equation}
We define the ``holonomy'' associated with the ``transverse'' 
connection $\bar{A}_\varphi$ at a vertex
$v$.  We will use this to construct the Hamiltonian. The definition is,
\begin{equation}
  h_\varphi(v,\rho) \equiv \exp\left( \frac{i}{2} 
\rho \bar{A}_{\varphi}(v)\right) =
\exp\left(i\rho \gamma K_\varphi(v)\right),
\end{equation}
one has that,
\begin{equation}
\hat{h}_\varphi(v_i,\rho) | g, \vec{\mu}> = | g,\mu_{v_1},
\ldots,\mu_{v_i}+\rho,\ldots>,
  \end{equation}
and in the case of $v_i$ not belonging to the initial spin network,
the action adds a vertex with $\mu=\rho$. The co-triad is
\begin{equation}
  \hat{E}^\varphi_m |g,\vec{\mu}> = 
\sum_{v\in V(g)} \mu_v \gamma \ell_{\rm Planck}^2
\delta_{m,n(v)}|g,\vec{\mu}>.
\end{equation}

We will now define the inverse operators that arise in the Hamiltonian. It should be
noted that similarly with what happens in loop quantum cosmology, the inverse 
operators are bounded. We will start by computing classically 
${\rm sgn}(E^\varphi)/\sqrt{|E^\varphi|}$. To do it we  note that,
\begin{eqnarray}
D_m &\equiv& 
\cos\left(\frac{\rho\gamma K_{\varphi,m}}{2}\right)
\left\{\sqrt{|E^\varphi_m|},\sin\left(\frac{\rho \gamma K_{\varphi,m}}{2}\right)\right\}
-
\sin\left(\frac{\rho\gamma K_{\varphi,m}}{2}\right)
\left\{\sqrt{|E^\varphi_m|},\cos\left(\frac{\rho \gamma K_{\varphi,m}}{2}\right)\right\}\\
&=&-\frac{G {\rm sgn}(E^\varphi_m) \gamma \rho}{4 \sqrt{|E^\varphi_m|}} 
\cos^2\left(\frac{\gamma\rho K_{\varphi,m}}{2}\right)
-\frac{G {\rm sgn}(E^\varphi_m) \gamma \rho}{4 \sqrt{|E^\varphi_m|}} 
\sin^2\left(\frac{\gamma\rho K_{\varphi,m}}{2}\right)\\
&=&-\frac{G {\rm sgn}(E^\varphi_m)\gamma\rho}{4 \sqrt{|E^\varphi_m|}},
\end{eqnarray}
which leads to 
\begin{equation}
\frac{{\rm sgn}(E^\varphi_m)}{\sqrt{|E^\varphi_m|}} = 
-\frac{4}{\gamma G\rho} D_m.
\end{equation}

Quantum mechanically one has
\begin{eqnarray}
\widehat{\frac{{\rm sgn}(E^\varphi_m)}{\sqrt{E^\varphi_m}}} |g,\vec{\mu}> &=&
\frac{4 i}{\gamma\ell_{\rm Planck}^2 \rho}
\left(
\cos\left(\frac{\gamma\rho \hat{K}_{\varphi,m}}{2}\right)
\left[\sqrt{|\hat{E}^\varphi_m|},
\sin\left(\frac{\gamma\rho \hat{K}_{\varphi,m}}{2}\right)\right]\right.\\
&&-\left. 
\sin\left(\frac{\gamma\rho \hat{K}_{\varphi,m}}{2}\right)
\left[\sqrt{|\hat{E}^\varphi_m|},
\cos\left(\frac{\gamma\rho \hat{K}_{\varphi,m}}{2}\right)\right]
\right)|g,\vec{\mu}>,
\end{eqnarray}
and noting that,
\begin{equation}
\sqrt{|\hat{E}^\varphi_m|} |g,\vec{\mu}> = 
\sum_{v\in V(g)} \delta_{m,n(v)} 
\sqrt{\gamma} \ell_{\rm Planck} |\mu_v|^\frac{1}{2} |g,\vec{\mu}>,
\end{equation}
one finally has,
\begin{equation}
\widehat{\frac{{\rm sgn}(E^\varphi_m)}{\sqrt{E^\varphi_m}}} |g,\vec{\mu}> =
\frac{2}{\sqrt{\gamma}\ell_{\rm Planck}\rho} \sum_{v\in V(g)} \delta_{m,n(v)} 
\left(|\mu_v+\frac{\rho}{2}|^\frac{1}{2}-
|\mu_v-\frac{\rho}{2}|^\frac{1}{2}\right)|g,\vec{\mu}>.
\end{equation}

The maximum value of this quantity occurs for $\mu_v=\rho/2$ and is
given by $2/(\sqrt{\gamma}\ell_{\rm Planck})$. For $\mu\gg \rho$ one
should recover the classical approximation. In that limit
$|g,\vec{\mu}>$ is an eigenstate with eigenvalue
$1/(\sqrt{\gamma}\ell_{\rm Planck} \sqrt{|\mu_v|})$ and satisfies that,
\begin{equation}
\hat{|E^\varphi|} \widehat{\left(\frac{{\rm sgn}(E^\varphi)}{\sqrt{|E^\varphi|}}\right)^2}|g,\vec{\mu}\rangle=|g,\vec{\mu}\rangle,
\quad \mu>>\rho
\end{equation}
which is the usual relation between the classical variables. In what follows
we will either work in the loop representation or the connection representation
as needed.

\subsection{The Hamiltonian constraint, Abelianization and 
traditional Dirac quantization}

Here we will show that one can Abelianize the Hamiltonian constraint,
discretize it and then quantize the discrete theory using the traditional
Dirac quantization procedure. It is known that for Abelian constraints 
the uniform discretization procedure coincides with the Dirac quantization
so we will not work it out explicitly. This quantization will provide
a baseline against which to compare quantizations in which we do not
Abelianize the constraint. The latter are more realistic since in 
the general theory it is unlikely that one will be able to Abelianize
the constraints. 

We start from
the classical expression (\ref{hamilclass}) for the Hamiltonian constraint,
\begin{equation}
  H=-{E^\varphi \over (x+a) \gamma^2}\left({\bar{A}^2_\varphi (x+a)\over 8}\right)'
-{E^\varphi \over 2 (x+a)}+ {3 (x+a) \over 2 E^\varphi} 
+ (x+a)^2 \left({1 \over E^\varphi}\right)'=0.
\end{equation}
We then Abelianize the  Hamiltonian (\ref{hamilclass}),
multiplying by $\frac{2(x+a)}{E^\varphi}$ and grouping terms as
\begin{equation}\label{abelianized}
H= \left(\frac{(x+a)^3}{(E^\varphi)^2}\right)'-1 -\frac{1}{4 \gamma^2}
\left((x+a) \bar{A}_\varphi^2\right)'.
\end{equation}

We wish to write the discretization in terms of classical quantities
that are straightforward to represent in the quantum theory. Here one
has to make choices, since there are infinitely many ways of
discretizing a classical expression. In particular, we will notice
that there exists, for this model, a way of discretizing the
constraint in such a way that it remains first class (more precisely,
Abelian) upon discretization. This is unusual, and we do not expect
such a behavior in more general models.

We now proceed to discretize this expression and to ``holonomize'' it,
that is, to cast it in terms of quantities that are easily representable
by holonomies,
\begin{equation}
  H^\rho_m = \frac{1}{\epsilon}\left[
\left(\frac{(x_m+a)^3 \epsilon^2}{(E^\varphi_m)^2}-
\frac{(x_{m-1}+a)^3\epsilon^2}{(E^\varphi_{m-1})^2}\right)
-\epsilon 
-\frac{1}{4 \gamma^2 \rho^2}
\left(
(x_m+a)\sin^2\left(\rho \bar{A}_{\varphi,m}\right)
-(x_{m-1}+a)\sin^2\left(\rho \bar{A}_{\varphi,m-1}\right)
\right)
\right],
\end{equation}
expression that recovers (\ref{abelianized}) in the limit $\epsilon\to
0$, $\rho \to 0$. From now on we will assume the spacing is uniform so
we are dropping the $m$ dependence of $\epsilon$. This expression is
immediately Abelian since it can be written as the difference of two
terms, one dependent on the variables at $m$ and the other at
$m-1$. Therefore each term has automatically vanishing Poisson
brackets with itself and with the other,
\begin{equation}
  H^\rho_m = \frac{1}{\epsilon} \left(\phi^\rho(x_m,E^\varphi_m,\bar{A}_{\varphi,m})
-\phi^\rho(x_{m-1},E^\varphi_{m-1},\bar{A}_{\varphi,m-1})\right),
\end{equation}
with 
\begin{equation}
  \phi^\rho(x_m,E^\varphi_m,\bar{A}_{\varphi,m}) = 
\frac{(x_m+a)^3}{(E^\varphi_m)^2}\epsilon^2 -x_m
-\frac{1}{4\gamma^2\rho^2}(x_m+a)\sin^2\left(\rho \bar{A}_{\varphi,m}\right).
\end{equation}
At the horizon, the boundary condition is $\phi^\rho(0,0,0)=0$. The
condition that the constraints vanish $H^\rho_m=0$ is equivalent to
$\phi^\rho(x_m,E^\varphi_m,\bar{A}_{\varphi,m})=0$.

When one has an Abelian set of constraints the uniform discretization
quantization procedure is equivalent to the Dirac quantization
procedure, as was shown in \cite{uniformprd}. So for simplicity we will
just follow the Dirac procedure. To implement the constraints as quantum
operators as one does in the Dirac procedure, it is convenient to solve
the constraints for the $E^\varphi_m$, 
\begin{equation}\label{convenient}
  E^\varphi_m = \pm \frac{(x_m+a)\epsilon}{
\sqrt{1-\frac{a}{x_m+a}+\frac{1}{4\gamma^2\rho^2}
\sin^2\left(2 \rho \gamma K_{\varphi,m}\right)}},
\end{equation}
and this relation can be immediately implemented as an operatorial
relation and find the states that satisfy it. It should be noted that
this relation can be implemented for other gauges as well in a
straightforward manner. The states are given by,
\begin{equation}\label{109}
  \Psi[K_{\varphi,m},\tau,a] =C(\tau,a) \exp\left(
\pm \frac{i}{\ell_{\rm Planck}^2}\sum_m
f[K_{\varphi,m}]\right), 
\end{equation}
where $C(\tau,a)$ is a function of the variables at the boundary $\tau$ and $a$, which has to solve the constraint at the boundary (we will discuss this later).
The functional $f$ has the form,
\begin{eqnarray}
  f[K_{\varphi,m}]&=& 
\frac{1}{4\gamma^2\rho^2\left(1-\frac{a}{x_m+a}\right)}(x_m+a)\epsilon
\left[F\left(\sin(2\gamma \rho K_{\varphi,m}),
\frac{i}{4\gamma^2\rho^2\left(1-\frac{a}{x_m+a}\right)}\right)\right.\\
&& \left.+ 2 F\left(1,
\frac{i}{4\gamma^2\rho^2\left(1-\frac{a}{x_m+a}\right)}\right)
{\rm sgn}\left(\sin(2\gamma\rho K_{\varphi,m})\right)\right] \nonumber
\end{eqnarray}
with $F(\phi,m)\equiv \int_0^\phi (1-m^2\sin^2 t)^{-1/2}dt$ the Jacobi
Elliptic function of the first kind. Notice that the continuum limit
of this expression for the state is immediate, i.e. the sum in $m$ 
becomes an integral.

We now need to impose the constraints on the boundary, in particular
$p^\tau = -a/2$ (in the limit $N\to\infty$). Quantum mechanically
$\hat{p}^\tau = -i \ell_{\rm Planck}^2 {\partial/\partial \tau}$ and
therefore,
\begin{equation}
  C(\tau, a) = C_0(a) \exp\left(-\frac{i a \tau}{2 \ell_{\rm Planck}^2}\right)
\end{equation}
and $C_0(a)$ is an arbitrary function. This is analogous to the
quantization that Kucha\v{r} found where one had wavefunctions that
only depended on the mass. We have therefore completely solved the
theory.

Remarkably, the physical state we found is normalizable in a kinematic
space of wavefunctions associated with one superselection sector, that
is, the space of $K_\varphi's$ defined in the interval $[0,\pi/(\gamma\rho)]$
(this defines an inner product in the bulk, the picture is easily 
completed in the boundary by considering functions $C_0(a)$ which are
square integrable, as we will do later).
In such space one can define operators associated to quantities that are
not observables in the theory. This is therefore well suited to treat
the problem of time in a relational way, as proposed by Page and
Wootters \cite{PaWo}. The main objection to the Page--Wootters construction
was that one could not construct conditional probabilities based on states
of the kinematical Hilbert space because the physical states were not
contained in the kinematical Hilbert space and therefore the conditional
probabilities were not well defined on physical states (see Kucha\v{r} 
\cite{kuchartime} for
a detailed discussion). Here, since the physical states exist in the
kinematical space one does not face that problem. Although in the
particular example we are considering there is not much point in
defining a relational time, given its simple and static nature, it
should be noted that the above result exhibits some level of
robustness. For instance, it is true every time we are considering
models where the variables are handled via the Bohr
compactification. This is true, for instance, in cosmological
models. It is also true if one couples the present model to a scalar
field. It may also hold in more complicated models when one gauge
fixes the diffeomorphisms. This is worthwhile further investigation.

The above quantization has taken place in the ``connection
representation''. We would like to see that one obtains equivalent
results in the loop representation.
It is more convenient to write (\ref{convenient}) operatorially as 
$\hat{O}_m\Psi=\Psi$ 
where,
\begin{equation}
  \hat{O}_m = \left(
\sqrt{1-\frac{a}{x_m+a}+\frac{1}{4\gamma^2\rho^2}
\sin^2\left(2 \rho \gamma \hat{K}^\varphi_m\right)} 
\frac{\hat{E^\varphi_m}}{(x_m+a)\epsilon}\right)^2
\end{equation}
since this expression is straightforward to represent in the 
loop representation,
\begin{eqnarray}
 && \left(1 -\frac{a}{x_m+a}+\frac{1}{8\gamma^2 \rho^2}\right)
\frac{\mu_m^2 \gamma^2 \ell_{\rm Planck}^4}{(x_m+a)^2\epsilon^2} 
\Psi(\mu_m)-
\left[ \frac{\mu_m^2 \gamma^2 \ell_{\rm Planck}^4}{(x_m+a)^2 \epsilon^2}
+ \frac{2 \mu_m \gamma^2 \ell_{\rm Planck}^4}{(x_m+a)^2 \epsilon^2}\right]
\frac{\Psi(\mu_m+4\rho)}{16\gamma^2 \rho^2}\nonumber\\
&&
-\left[ \frac{\mu_m^2 \gamma^2 \ell_{\rm Planck}^4}{(x_m+a)^2 \epsilon^2}
- \frac{2 \mu_m \gamma^2 \ell_{\rm Planck}^4}{(x_m+a)^2 \epsilon^2}\right]
\frac{\Psi(\mu_m-4\rho)}{16\gamma^2 \rho^2}=\Psi(\mu_m).
\end{eqnarray}

The above equation is a recursion relation that implies that in
the functions $\Psi(\mu_m)$ the possible values of the $\mu$'s are
$\mu = \mu(r)= \pm r+4 n \rho$ with $r\in [0,2\rho]$ and $n$ an
integer. Solutions for different values of $r$ are therefore not
connected, and there is therefore a superselection. It is suggestive
to compare this expression with the one obtained in loop quantum
cosmology, where we see that for each point in the radial direction
our wavefunctions have a similar recursion relation to those in loop
quantum cosmology for the wavefunction of the universe.

One can also see that the solution to this recursion relation can be
obtained via the ``loop transform'' from the solution in the connection
representation we found. Each solution in the connection representation
corresponds to a given r-sector of the superselection rule via,
\begin{equation}
  \Psi_r(\mu_m) =\frac{\rho \gamma}{\pi}\int_0^{\pi/(\rho \gamma)} d K_\varphi
\Psi(K_\varphi) \exp\left(2\rho K_\varphi \gamma \mu_m(r)\right). 
\end{equation}

The physical space of states is a Hilbert space with a natural inner
product given by the $L^2$ norm in the variable $a$, and one must
demand that the functions $C_0(a)$ be square integrable.  The theory
has two independent observables, the mass and its canonically
conjugate variable. As mentioned above, the wavefunctions are
functions of the mass. One can define observables with more geometric
content, for instance the metric in a given gauge, as a function of
the mass, using for example equations
(\ref{metricgauge1}-\ref{metricgauge3}), and similarly in
Schwarzschild coordinates.  The continuum limit is taken trivially,
since the variables in a given coordinate system are uniquely defined
in terms of the functions of the physical space and one approximate a
continuum solution as accurately as one needs by reducing the
stepsize.  If one keeps the theory discrete, the relation is only
approximate.  It should be noted that the continuum limit is achieved
in the limit $\epsilon\to 0$ and $\rho\to 0$. If one adopts the point
of view commonly used in loop quantum cosmology, that the quantum of
distance should have a minimum value, then one would not expect to
take the limits $\rho$ and $\epsilon$ 
going to zero, but to keep the parameters at a
minimum value. In such a case one could expect to eliminate the
singularity.  This is plausible since then the triads would likely not
go to zero.  However, this would require a more careful analysis using
coordinates that actually reach the singularity, which is not the case
for the coordinates we have chosen so at the moment we cannot
conclusively state what happens at the singularity. Also, it should be
noted that in the most recent treatments of loop quantum cosmology the
quantity that plays the same role as $\rho$ is not taken to have a
constant value, but to depend on the dynamical variables
\cite{asreview}. If one followed a similar approach here, the results
we derived would change. The results would remain valid in the
asymptotic region far away from the horizon, since there $\rho$ would
tend to a constant.

Another interesting aspect is that although we regularized the
theory using a lattice, the level of ambiguity  of
the construction is limited, in part in order to have the
constraints be Abelian. Although very limited, this example suggests
that the use of a lattice regularization is not necessarily fraught
with uncontrollable ambiguities.

An aspect we have not emphasized is that in the calculations we have
assumed that $a\ge 0$. Since $a$ is the dynamical variable on which
the wavefunctions end up depending, it is problematic to work with an
operator with a continuum spectrum and positive eigenvalues only.  A
better way to handle this is to allow $a$ to take all possible real
values and then the gauge fixing condition we introduced should read
$E^x=(x+|a|)^2$, and all subsequent equations involving $x+a$ have to
be modified accordingly (the conjugate momentum should be modified 
as well). In this treatment the mass is given by
$|a|/2$ and is automatically positive. It is interesting to speculate
what would happen if one wished to consider negative masses. What
is clear at this stage is that the analysis would differ significantly
with what we did in this paper, since in the Cauchy surface one would
have to take into account the singularity and one cannot limit oneself
to study the ``exterior'' as in the case of positive mass, where there
is a well defined causal boundary at the horizon. The discussion of
negative masses is better postponed till one can handle the interior
problem and discuss the possibility of eliminating the singularity,
since in the negative mass case the singularity has to be faced
from the outset.

\section{Quantization using uniform discretizations}

Although we have succeeded in quantizing the model using the traditional
Dirac quantization, we saw that in order to do this one had to use
a property ---the Abelianization of the Hamiltonian constraint--- that
is unlikely to hold in more general models. It is therefore worthwhile
asking: what would have happened if we didn't Abelianize the constraint?
In such case the traditional Dirac quantization would not have 
succeeded and we would have to resort to alternative proposals, like
the uniform discretizations.  Generically
the discrete constraints that are obtained by discretizing a field
theory with first class constraints are second class and become first
class only in the continuum limit. We would like to develop a
quantization strategy for such systems. An immediate answer to the
problem is to deal with the second class constraints using the Dirac
procedure. This is likely to be very onerous in cases of interest, and
it does not take advantage of the fact that in the continuum limit the
constraints become first class.

Here is where our {\em uniform discretization} approach can help.  In
this approach the discrete theory has no constraints and nevertheless
is capable of approximating the continuum theory in a controlled
fashion. What we would advocate is to construct the discrete theory
using the uniform discretization approach and then proceed to quantize
the resulting discrete theory. In some cases one will be able to take
the continuum limit in the quantum theory and this completes the
quantization of the original continuum theory one started with
satisfactorily.  In some cases, as it is likely to be the case in the
most interesting situations, it might occur that the continuum limit
cannot be taken in the quantum theory. In such cases the approach we
advocate is the following: what matters in a quantization procedure is
to recover in the semiclassical limit the classical theory one started
from plus corrections. We know the classical discrete theory we
constructed approximates the continuum theory well. We therefore
expect the quantum discrete theory to also approximate well the
quantum continuum theory, even in cases where we cannot construct the
latter exactly via this method, at least for certain states (it will
obviously fail, for instance, for states that probe lengths smaller
than the lattice spacing). We would therefore end with a quantum
theory that approximates semiclassically the classical theory we
started with, plus corrections, i.e. the goal we were trying to
achieve.

Since these points of view imply a radical departure from the
traditional Dirac method, it is worthwhile investigating how they
perform in the face of concrete models where one can carry out the
computations explicitly. In this section we would like to apply the
uniform discretization approach to the treatment of spherically
symmetric midi-superspaces.  We will see that the approach is
feasible but calculations can become quite involved if one
does not choose to Abelianize the constraints.

\subsection{A brief summary of Uniform discretizations}

To recap briefly on previous discussions of uniform discretizations
we would like to summarize its application to a field theory.
We assume one is starting with a theory with variables $q^i(x,t)$. 
One discretizes the underlying space-time manifold. The action can
therefore be approximated by,
\begin{eqnarray}
S&=& \sum_n L(n,n+1) \\
&&L(n,n+1) \equiv  \sum_m \Delta\epsilon L([q^i_{m,n}],[q^i_{m,n+1}]),
\end{eqnarray}
where $q^i_{m,n}$ represents a discretization of $q^i(x,t)$ with 
$m$ representing the array points in the spatial discretization and
$n$ the ones in the timelike direction. 
The notation $[q^i_{m,n}]$ means the variable $q^i_{m,n}$ and
some neighboring $q^i$'s determined by the scheme used to discretize
spatial derivatives. For instance, if one uses a non-centered first
order scheme one would have $[q^i_{m,n}]=(q^i_{m,n}, q^i_{m+1,n})$.
$\epsilon$ is the coordinate spatial separation between neighboring
points in the lattice and $\Delta$ the time-like separation of
neighboring points.  Though we do not make it explicit, $L$ depends on
both $\epsilon$ and $\Delta$. We choose a first order approximation
for the time derivatives since it simplifies constructing a canonical
theory (for treatments with more than two time levels see
\cite{baleanu}). One works out the canonical momenta for the discrete
action,
\begin{equation}
p^k_{m,n+1}\equiv \frac{\partial L(n,n+1)}{\partial q^k_{m,n+1}}.
\end{equation}
One can also work out the momenta at $n$, which via the Lagrange 
equation,
\begin{equation}
\frac{\partial L(n,n+1)}{\partial q^i_{m,n}} + 
\frac{\partial L(n-1,n)}{\partial q^i_{m,n}} =0,
\end{equation}
 yields,
\begin{equation}
p^k_{m,n}\equiv -\frac{\partial L(n,n+1)}{\partial q^k_{m,n}}.
\end{equation}
These equations imply a canonical transformation between $q,p$ at
level $n$ and level $n+1$ via a type I generating function given by
$F(q_n,q_{n+1})=-L(n,n+1)$.  It should be noted that we have not made
a distinction between variables and Lagrange multipliers.  Upon
discretization, variables that in the continuum were Lagrange
multipliers become evolution variables and they get determined by the
evolution equations (this is a generic statement for Lagrange
multipliers associated with diffeomorphism invariances, for many
details concerning the Dirac treatment of discrete systems and
particular examples see
\cite{Dirac}). Equations that in the continuum were constraint
equations now become evolution equations.  The number of degrees of
freedom is therefore larger than in the continuum theory. There will
be different solutions in the discrete theory that correspond to
different approximations or parameterizations of the same solution in
the continuum theory. The determination of the Lagrange multipliers
has proved problematic in previous examples we have studied. The
resulting equations are usually polynomials of high order that can
have complex solutions or develop discontinuities and branches.  This
may translate in that the discrete theory produces solutions that are
not close to the constraint surface of the continuum theory throughout
the whole evolution, and may depart significantly from it at certain
points.  This was usually considered a major obstacle to the use of
these discrete theories. The uniform discretization approach bypasses
this problem.

The approach consists in replacing the evolution equations
that we obtained for the dynamical variables of the problem
by a set of evolution equations that in a sense we will make
precise later on preserves the constraints of the continuum
theory at a given level of approximation. The reason we can
do this is that there is a large amount of freedom in 
how one discretizes a theory. In a sense we will be exploiting
this freedom to our advantage.

To construct the uniform discretizations we note that
since the evolution is given by a canonical transformation, 
generically one can find an infinitesimal generator for it.
We will call this generator $\mathbb{H}$. In terms of this generator,
the discrete evolution of the dynamical variables can be written as,
\begin{equation}
q^i_{m,n+1} = e^{\{\bullet, \mathbb{H}\}}\left(q^i_{m,n}\right) \equiv
q^i_{m,n}+ \{q^i_{m,n},\mathbb{H}\}+
\frac{1}{2}\{\{q^i_{m,n},\mathbb{H}\},\mathbb{H}\}+\cdots
\end{equation}

We will assume a discretization has been chosen such that the form of
$\mathbb{H}$ is given as
$H(q,p)=f(\phi_1(q,p),\ldots,\phi_P(q,p))$ where
$\phi_1,\ldots,\phi_P$ are discretizations of the constraints of the
continuum theory. To simplify the notation we have dropped the fact
that the discretizations of the constraints will generically be
functions of $[q^i_{m,n}],[p^i_{m,n}]$. The function 
$f$ has the following properties:
a) $f(x_1,\ldots,x_P)=0 \iff x_a=0, a=1\ldots P$ and otherwise
$f>0$; b) $\frac{\partial f(0,\ldots,0)}{\partial x_b}=0$; c) ${\rm
det} \frac{\partial^2 f}{\partial x_a\partial x_b}\neq 0,\forall
x$ and d) $f(\phi_1(q,p),\ldots,\phi_P(q,p))$ is defined for
all $(q,p)$ in the complete phase space.

Discretizations of this sort can be constructed by suitable choices
when one discretizes the action, as we proceeded. Further discussion
on this is in reference \cite{uniformprl}. However, we will not use in
any way that the discretization was arrived to in this form so we
refer the reader to the reference and just take the discretization as
a given.

A particularly simple example of such a function, that is useful in many
systems is to just consider the sum of the squares of the discretized
constraints.  It is immediate that the evolution equations we wrote
preserve exactly $\mathbb{H}$. Therefore this means that if one chooses
a small initial value for $\mathbb{H}$ one is guaranteeing that the sum
square of the constraints is kept small and therefore one is not 
departing significantly from the constraint surface. This is also the
case if one takes a more general $f$ as well.

Let us now show that this evolution does indeed have the correct
continuum limit. For concreteness, and since this is general enough to
accommodate the example we study in this paper, we consider a system
that represents a discretization of a $1+1$ dimensional field theory
of the continuum with one first class constraint. We assume the
spatial direction has a finite extension $[x_0,x_1]$. We have a
lattice with $N$ points such that $|x_1-x_0|=N\epsilon$. The time-like
direction starts at $t_0$ and has spacing $\Delta$. The field has $R$
components $q^i(x), i=1\ldots R$ and canonical momenta $p^i(x)$. One
has that,
\begin{eqnarray}
q^i_{m,n} &=& q^i(x_0+m\epsilon,t_0+n\Delta),\label{corres1}\\
\frac{p^i_{m,n}}{\epsilon} &=& p^i(x_0+m\epsilon,t_0+n\Delta),\\
x&=&\lim_{\epsilon \to 0} x_0 +m \epsilon,\\
t&=&\lim_{\Delta \to 0} x_0 +m \Delta,\\
q^i(x,t)&=& \lim_{\epsilon,\Delta\to 0} q^i(x_0+m\epsilon,t_0+n\Delta),\\
p^i(x,t) &=& \lim_{\epsilon,\Delta \to 0} p^i(x_0+m\epsilon,t_0+n\Delta).\label{corresn}
\end{eqnarray}
We also assume a single (field-theoretic) constraint in the continuum
that corresponds to $N$ constraints
$\phi_a([p_{n,m}],[q_{n,m}],\epsilon)$ in the discrete theory. The
generator is given by 
$\mathbb{H} = k^2 \sum_{a=1}^N
\phi_a^2/2$. In units where $\hbar=c=1$ the constraints have dimensions of 
length$^{-1}$ and the constant $k$ has dimensions of length so 
that the generator is dimensionless.
We assume that the
constraints have been appropriately rescaled such that,
\begin{equation}
\lim_{\epsilon\to 0} \mathbb{H} = \frac{k^2}{2} \int dx \phi^2[q(x),p(x)], 
\end{equation}
where we emphasize that the constraint has a functional dependence on $p,q$.

Since $\mathbb{H}$ is preserved upon evolution, we take $\mathbb{H}=\delta^2/2$ with
$\delta$ a constant. We define $N_a\equiv
k \phi_a([q],[p])/\delta$ which satisfies $\sum_{a=1}^N N_{a}^2=1$.  We note
that,
\begin{equation}
\frac{q^i_{m,n+1}-q^i_{m,n}}
{k \delta} = \frac{1}{k\delta}\{q^i_{m,n},\mathbb{H}\} +O(\Delta)= \sum_{a=1}^N
\{q^i_{m,n},\phi_a([q],[p])\} N_a+O(\Delta)
\end{equation}
and identifying $\Delta=k \delta$ and 
taking the limit ${\epsilon\to 0}$ (and therefore
$N\to \infty$), ${\delta \to 0}$  one has that,
\begin{equation}
\dot{q}^i(x,t) = \int dz \{q^i(x,t), \phi[q(z,t),p(z,t)]\} N_z
\end{equation}
where we have taken the limit of vanishing time and spatial spacings
and therefore the number of spatial points and of constraints to
infinity, as one would expect in a field theory.   In the discrete
theory the canonical Poisson brackets are
$\{q^i_{m,n},p^j_{m,n}\}=\delta^{i,j} \delta_{m,n}$ with Kronecker
deltas and in the continuum $\{q^i(x,t),p^j(z,t)\} =\delta^{ij}
\delta(x-z)$. We have assumed that the model has one spatial dimension,
as we consider in this paper, it is straightforward to
generalize the construction to more spatial dimensions. Although we 
did not specify the field theory we were considering, it should be 
noted that in diffeomorphism invariant theories, to construct the
generator one needs to integrate a density of weight one. Therefore
one should choose appropriately densitized versions of the constraints.

The above equations are valid if the constraints are first class in
the continuum theory and the corresponding quantities in the discrete
theory become first class constraints in the continuum
limit. Generically, constraints that are first class in the continuum
become second class quantities when discretized and will not
necessarily become first class in the continuum limit of the discrete
theory, which might also fail to exist altogether. To be precise, by
``second class'' in the discrete theory we mean that the vanishing of
the quantities does not imply that their Poisson bracket
vanishes. Notice that the equations specify the values of the Lagrange
multipliers, they are given, up to a factor, by the values of the
constraints. This therefore mandates that the constraints must be
first class since only then one is free, in the continuum theory, to
pick a particular set of Lagrange multipliers, which is what our
method is doing.  If the constraints are second class, the Lagrange
multipliers are determined by the Dirac procedure and therefore the
above procedure can be potentially inconsistent. There are two ways to
proceed in this case. One of them is to use Dirac brackets in all the
above expressions.  Then the method still works as specified. Another
possibility is to pretend the constraints are first class and then to
study the continuum limit of the theory. If the continuum limit exists
and in such limit the constraints become first class, then one may be
able to use the discrete theory to construct the continuum limit and
use it as a vehicle for quantization. Although it might appear that
this latter point of view is less warranted than the first, in
practice for cases of interest (as in general relativity in $3+1$
dimensions) the use of the Dirac brackets will lead to unsolvable
equations and one may opt for the second avenue as a route for
quantization.  In fact, the latter point of view is likely to become
the preferred one for cases of interest. The reason for this is that
even in the case in which the continuum limit does not exist, one can
use the constructed discrete theory as an approximation of the
continuum one since the continuum constraints can be kept small. There
might be other reasons why one does not want to consider the continuum
limit. For instance, as occurs in loop quantum gravity, the
kinematical space is not well suited for taking spatially continuum
limits since one expects space to be quantized, as indicated by the
quantization of the area, volume, etc. We will see how this operates
in the example we discuss in this paper.

\subsection{Spherically symmetric vacuum gravity via uniform discretizations}

We would like to treat the model using uniform discretizations. Here
one faces two possibilities. Since the discrete constraints are not
first class, one can only ensure that the uniform discretizations will
have the continuum limit if one treats them using Dirac brackets.  However,
one does not expect that in more complicated models one will be able
to take this path. We will therefore choose to treat the model using a
technique more suited for more general models. The technique is to
implement the uniform discretizations without using the Dirac
brackets. Since one knows that in the continuum limit the constraints
become first class, it is possible that in the continuum limit one
will recover the desired theory. We will actually use some of the
ambiguities in the discretization to make this outcome more likely.

We will proceed to construct the quantity $\mathbb{H}$ which
corresponds to the square of the Hamiltonian constraint. Here one
faces a significant degree of ambiguity in choosing how to discretize
the expression. To guide us in this we would like to require some
conditions on the discretization that make it more likely that the
continuum quantum limit will be achieved. The first condition is
that when we take the continuum limit in the classical theory, the
classical continuum constraint algebra should be reproduced. That is,
the continuum limit of the Poisson bracket of the constraints of the
discrete theory should reproduce the continuum classical constraint
algebra. The second condition is that the spectrum of the quantity
$\mathbb{H}$ contain the zero eigenvalue in the combined classical and
continuum limit. If these conditions are met, then one can be satisfied
that one constructed a suitable quantum theory in the continuum limit.

In achieving the above limit one expects two different types of
problems.  On the one hand, it could be that the constraints have been
discretized in such a way that even classically they do not reproduce
the constraint algebra in the continuum limit and this may imply that
even classically $\mathbb{H}$ does not vanish in the continuum
limit. In addition to this problem, one may have quantum
anomalies. That is, even if one ensured that the algebra was
reproduced classically in the continuum limit, upon quantization one
may fail to reproduce the algebra in the continuum limit. The two
types of problems can be characterized by the appearance of terms of
order $\epsilon^n$ with $\epsilon$ the spatial separation and $n$ some
power in the case of classical effects and the appearance of terms
$(\ell_{\rm Planck}/\epsilon)^m$ with $m$ some power in the case of
the quantum anomalies. For things to work out in the limit we should
have that $n>0$ ($m\ge0$ due to how quantum anomalies
arise). This suggests that the classical continuum limit should be 
taken $\ell_{\rm Planck}\to 0$ first and then $\epsilon\to 0$. 

So what we are arguing is that due to the presence of quantum
anomalies one could face the situation where there is no well defined
quantum continuum theory if $m>0$, and nevertheless the discrete
quantum theory is able to produce an acceptable continuum classical
limit. This has important parallels with proposals by Klauder to
extend the more traditional approaches to quantization \cite{klauder2},
and proposals on how to handle Thiemann's master constraint in the
case in which it does not have zero eigenvalue
\cite{dittrichthiemann}.

We already have a candidate for the constraints that satisfies all
the above conditions: the Abelianized version. What we would like
to do, however, is to consider other versions of the constraint,
where the above problems arise, and show that one can construct
a suitable quantization. At this point one has to choose a discretization
and a quantization of $\mathbb{H}$ that satisfies the two conditions
we outlined above. There is no generic algorithmic procedure to
construct such a discretization and quantization. In a generic
situation one can imagine proposing parameterized discretizations
and checking that the conditions are satisfied by choosing 
parameters. It is also possible to add terms that vanish in the
continuum limit to the constraints to achieve the desired 
result. In the particular model we are studying we will construct
a suitable discretization by starting from the Abelian version of
the Hamiltonian constraint (\ref{abelianized}), and we multiply
it times powers of the triad so one gets a non-Abelian constraint.
In this case we will choose to multiply it times a fourth power
of the triad since this simplifies quantizing it in the loop
representation. The classical discrete Hamiltonian constraint we
take is,
\begin{eqnarray}
  H^\rho_m &=& 
\frac{1}{\epsilon^5 (x_m+a)^2(x_{m-1}+a)^2}
\left[
(E^\varphi_{m-1})^2 \left((x_m+a)^3+\epsilon^3\right)\epsilon^2
-
(E^\varphi_{m})^2 \left((x_{m-1}+a)^3+\epsilon^3\right)\epsilon^2
\right.\nonumber\\
&&\left.
-
(E^\varphi_{m})^2 (E^\varphi_{m-1})^2 
\left(\epsilon+ 
\frac{(x_m+a)\sin^2\left(2\rho \gamma K_{\varphi,m}\right)
-(x_{m-1}+a)\sin^2\left(2\rho \gamma K_{\varphi,(m-1)}\right)}
{4\gamma^2 \rho^2}\right)\right].
\end{eqnarray}

The quantization we choose, in order for the operator to be Hermitian
is,
\begin{eqnarray}
 \hat{H}^\rho_m &=& 
\frac{1}{\epsilon^5 (x_m+a)^2(x_{m-1}+a)^2}
\left[
(\hat{E}^\varphi_{m-1})^2 \left((x_m+a)^3+\epsilon^3\right)\epsilon^2
-
(\hat{E}^\varphi_{m})^2 \left((x_{m-1}+a)^3+\epsilon^3\right)\epsilon^2
\right.\nonumber\\
&&-(\hat{E}^\varphi_{m})^2 (\hat{E}^\varphi_{m-1})^2 \epsilon
-
(\hat{E}^\varphi_{m})^2 (\hat{E}^\varphi_{m-1})^2 
\left(
\frac{(x_m+a)\sin^2\left(2\rho \gamma \hat{K}^\varphi_m\right)
-(x_{m-1}+a)\sin^2\left(2\rho \gamma \hat{K}^\varphi_{m-1}\right)}
{8\gamma^2 \rho^2}\right)
\nonumber\\
&&
-
\left.
\left(
\frac{(x_m+a)\sin^2\left(2\rho \gamma \hat{K}^\varphi_m\right)
-(x_{m-1}+a)\sin^2\left(2\rho \gamma \hat{K}^\varphi_{m-1}\right)}
{8\gamma^2 \rho^2}\right)(\hat{E}^\varphi_{m})^2 (\hat{E}^\varphi_{m-1})^2 
\right]
\end{eqnarray}

From here one can construct the operator $\hat{\mathbb H}=\sum_m
\left(\hat{H}^\rho_m\right)^2$, and we do not need to include powers
of the determinant of the metric, since one has fixed spatial
diffeomorphisms and therefore the spatial integral of the Hamiltonian
as a scalar is well defined. The task now is to solve the eigenvalue
problem and find the minimum eigenvalue of this positive definite
quantum mechanical operator. This is a well defined problem in quantum
mechanics, but it is far from trivial. Further studies on how to handle
these types of problems are  clearly needed. 

For the particular model at hand we can however determine an upper
bound on the value of the minimum eigenvalue. To do this we use the
states (\ref{109}), which are normalized in the kinematical inner product
\begin{equation}
\prod_m \frac{\gamma \rho}{\pi} \int_0^{\pi/(\gamma \rho)} dK_{\varphi,m}
\int_0^\infty da 
\Psi^*[K_{\varphi,m},\tau,a] 
\Psi[K_{\varphi,m},\tau,a] =1.
\end{equation}

We use this inner product to compute the expectation value of
$\hat{\mathbb{H}}$ for these states, and we get,
\begin{equation}
  \langle\Psi|\hat{\mathbb{H}}|\Psi\rangle= 
 C_1 \frac{\epsilon^3}{a^3} 
+C_2 \frac{\ell_{\rm Planck}^2}{a^3}
+C_3 \frac{\ell_{\rm Planck}^4}{a \epsilon^3}
+\sum_{n=3}^8 C_{n+1}
\frac{\ell_{\rm Planck}^{2n}}{a \epsilon^{n+1}},
\end{equation}
where the constants $C_i$ are of order unity. Since the inner product
has an integration over $a$ the above expressions in the powers of
$a$ really correspond to expectation values, for instance the first
term strictly speaking should be $C_1 \epsilon^3\langle a^{-3}\rangle$.  
The expectation value above
satisfies the conditions we imposed for the continuum classical limit,
i.e. it goes to zero as $\epsilon\to 0$, $\ell_{\rm Planck}\to 0$.
Moreover, in the continuum quantum limit ($\epsilon\to 0$ but
$\ell_{\rm Planck}$ finite) it is divergent. Notice that this problem
may arise even for the exact eigenvector corresponding to the minimum
eigenvalue of ${\mathbb H}$. That is, it is possible that one may not
be able to use the discrete quantum theory to define a continuum
quantum theory as a limit. The minimum value of bound for the
expectation value is achieved (assuming a large value of $\langle a\rangle$) 
for $\epsilon=\sqrt[3]{\ell_{\rm
Planck}^2 \langle a\rangle}$. For that value $\langle\hat{\mathbb{H}}\rangle \,
{\lower 3pt\hbox{$\mathchar"218$}}
\!\!\!\!\! \raise 2.0pt\hbox{$\mathchar"13C$}\,
\ell_{\rm Planck}^2/\langle a\rangle^2$. 
Just for reference, for a Solar sized black
hole, this amounts to $10^{-80}$.

\section{Conclusions}

We have analyzed spherically symmetric quantum gravity in the
connection and loop representations, outside the horizon. We were able
to find a gauge fixing of the spatial diffeomorphisms that yields a
theory with an Abelian Gauss law and a Hamiltonian constraint with a
first class algebra with structure functions. A redefinition of the
constraints yields an Abelian algebra of constraints. The latter can be
quantized using the traditional Dirac procedure and yields a quantum
theory in the connection and loop representation that coincides with
Kucha\v{r}'s treatment in terms of the traditional variables. We show
one can discretize the theory to regularize the Hamiltonian during
quantization in a reasonably unambiguous way and take the continuum
limit, which is well defined. 

An interesting point is that in the gauge fixed model the physical
quantum states of a given superselection sector are part of the
kinematical Hilbert space. This allows, for instance, to implement in
a well defined way a relational solution to the problem of time. 
This behavior is also expected in other models and needs to be 
further explored.

Although we succeeded in implemented a traditional Dirac
quantization for the model, this was done at a certain price.
On one hand we gauge fixed the diffeomorphisms. On the other
we Abelianized the Hamiltonian constraint through a rescaling.
Gauge fixing the diffeomorphisms may be viewed as a legitimate
approach to gain understanding into models, but ultimately is
faced with the usual question of if the quantization obtained
is equivalent to a quantization performed in another gauge. 
The Abelianization of the Hamiltonian is a property that as
far as we know, only holds in the gauge fixed model and 
appears unlikely to hold in other models generically (although
an Abelianization without gauge fixing was achieved using
the complex form of the Ashtekar variables in \cite
{thiemannthesis}).

In view of the mentioned limitations of the traditional approach,
we have also made a first exploration of the application of the
uniform discretization approach to the system when we do not
Abelianize the algebra of constraints.  We have not done a complete
analysis of such a model, since the calculations needed are
significant. We note that it is plausible that one may produce a quantum
theory that does not have a continuum limit, but has a classical
continuum limit. This behavior may also be expected 
in more complicated models, where constraints generically cannot be
Abelianized. Ambiguities in the discretization can be reduced by
choosing things in such a way that the quantum theory approximates the
continuum as much as possible. In this example, if one applied this
criterium, one would recover the Abelian theory previously mentioned.

Future developments will consist in studying the complete space-time
using Kruskal-type coordinates and to see what effect one would have
on the singularity due to the use of the loop representations. This
should also be compared with mini-superspace treatments of the
interior based on the isometry of the interior space-time with
Kantowski--Sachs \cite{ks}, although the latter should be more limited
since the isometry is limited to the interior region.  We see in our
treatment some initial hints of singularity removal since the discrete
equations resemble those found in loop quantum cosmology, though
further details need to be considered.

\section{Acknowledgements}

We wish to thank Abhay Ashtekar, Martin Bojowald, Parampreet Singh,
Thomas Thiemann and Madhavan Varadarajan for comments and two
anonymous referees for corrections.  We also wish to thank the Kavli
Institute for Theoretical Physics at the University of California at
Santa Barbara for hospitality.  This work was supported in part by
grant NSF-PHY-0554793, PHY-0551164, funds of the Hearne Institute for
Theoretical Physics, FQXi, CCT-LSU and Pedeciba (Uruguay).

\end{document}